\documentclass[12pt]{article}
\pdfoutput=1
\usepackage{geometry,enumerate,amsmath,amssymb}
\usepackage{fullpage}
\usepackage{graphicx}
\usepackage{bm}
\numberwithin{equation}{section}
\newcommand{\be}{\begin{equation}}
\newcommand{\ee}{\end{equation}}
\newcommand{\bea}{\begin{eqnarray}}
\newcommand{\eea}{\end{eqnarray}}
\renewcommand{\epsilon}{\varepsilon}

\usepackage{breqn}
\begin{document}
\title{
  A hyperbolic analogue of the Atiyah-Hitchin manifold
  }
\author{
  Paul Sutcliffe\\[10pt]
 {\em \normalsize Department of Mathematical Sciences,}\\
 {\em \normalsize Durham University, Durham DH1 3LE, United Kingdom.}\\ 
{\normalsize Email:  p.m.sutcliffe@durham.ac.uk}
}
\date{January 2022}

\maketitle
\begin{abstract}
The Atiyah-Hitchin manifold is the moduli space of parity inversion symmetric charge two $SU(2)$ monopoles in Euclidean space. Here a hyperbolic analogue is presented, by calculating the boundary metric on the moduli space of parity inversion symmetric charge two $SU(2)$ monopoles in hyperbolic space. The calculation of the metric is performed using a twistor description of the moduli space and the result is presented in terms of standard elliptic integrals. 
\end{abstract}

\newpage
\section{Introduction}\quad
Non-abelian magnetic monopoles are topological soliton solutions of Yang-Mills-Higgs gauge theories in three-dimensional space. In this paper the gauge group is taken to be $SU(2)$, with the Higgs field in the adjoint representation, and a vanishing Higgs potential. The number of monopoles is a positive integer $N$, referred to as the charge, since the magnetic charge of an $N$-monopole is equal to $N$ in suitable units. In the above system, magnetic monopoles are BPS solitons, with a moduli space of static $N$-monopole solutions that describes monopoles with arbitrary positions and internal relative phases.

Some time ago, Manton proposed that the dynamics of $N$ slowly moving monopoles could be described by geodesic motion in the $N$-monopole moduli space, equipped with a metric induced from the field theory kinetic energy \cite{Ma1}. This approach was later validated by rigorous mathematical analysis \cite{Stu}. To study the scattering of a pair of monopoles within this scheme requires knowledge of the metric on the 2-monopole moduli space. By fixing the centre of mass, the problem reduces to finding the metric on the 4-dimensional manifold of parity inversion symmetric 2-monopoles. This manifold is known as the Atiyah-Hitchin manifold, following their ingenious calculation of the metric by exploiting the fact that this is a hyperk\"ahler manifold \cite{AH}. Their study of the geodesics of this manifold revealed remarkable properties of monopole dynamics, including the right angle scattering of monopoles in a head-on collision: a phenomenon that is now ubiquitous in topological soliton systems.

Atiyah observed that many of the properties of monopoles in Euclidean space survive the transmutation to hyperbolic space \cite{At}. Furthermore, if the curvature of hyperbolic space is tuned to a specific value, determined by the magnitude of the Higgs field at spatial infinity, then hyperbolic monopoles correspond to circle-invariant Yang-Mills instantons in 4-dimensional Euclidean space, with the number of monopoles equal to the number of instantons. This simplifies a number of features in comparison to the case of generic constant negative curvature and will therefore be assumed from now on.

An important distinction between Euclidean and hyperbolic monopoles concerns their dynamics. The moduli space of static hyperbolic $N$-monopoles cannot be naturally equipped with the metric induced from the field theory kinetic energy, because the induced metric diverges. Braam and Austin defined an alternative metric on the hyperbolic $N$-monopole moduli space \cite{BA}, obtained as a boundary metric using the abelian connection of the hyperbolic monopole on the sphere at infinity, but its relation to monopole scattering was unclear. Recently, this metric has been placed in a more general context of boundary metrics on soliton moduli spaces and evidence presented to support the interpretation of the geodesics of such metrics in terms of soliton dynamics \cite{Su1}. This motivates the present work, to obtain a hyperbolic analogue of the Atiyah-Hitchin manifold by calculating the boundary metric on the moduli space of parity inversion symmetric charge two hyperbolic monopoles. The method exploits a description of the moduli space in terms of spectral curves in mini-twistor space, and yields an explicit expression for the metric in terms of standard elliptic integrals.

\section{Spectral curves of hyperbolic monopoles}\quad
The mini-twistor space of three-dimensional hyperbolic space is the space of its oriented geodesics. Each oriented geodesic may be labelled by a pair of Riemann sphere coordinates, $(\eta,\zeta)$, where the geodesic starts at $-1/\bar\eta$ and ends at $\zeta$, both regarded as points on the sphere at infinity in the unit ball model of hyperbolic space. The mini-twistor space is therefore
$\mathbb{CP}^1 \times\mathbb{CP}^1-\bar\Delta,$ where $\bar\Delta$ is the anti-diagonal, $\bar\eta\zeta=-1$, that must be removed because the start and end point of a geodesic are different.

There is a bijective correspondence between hyperbolic monopoles and spectral curves, which are algebraic curves in mini-twistor space satisfying certain reality and non-singularity conditions \cite{At}.
 Briefly, the spectral curve of an $N$-monopole 
 is a biholomorphic curve in mini-twistor space of bidegree $(N,N)$, so it
 may be written in the form
\be S(\eta,\zeta)= \sum_{i=0,j=0}^N c_{ij}\eta^i\zeta^j=0. 
\label{gensc}
\ee
The reality condition on the complex constants $c_{ij}$, 
that follows from reversing the orientation of the geodesic,
is
\be
\overline{c_{ij}}=(-1)^{N+i+j}c_{N-j,N-i}.
\label{reality}
\ee
The spectral curve describes all geodesics along which
a certain linear operator, constructed from the monopole fields, has a normalizable solution. This is equivalent to imposing non-singularity conditions on the algebraic curve that can be written in terms of relations between integrals of holomorphic differentials around particular cycles. Roughly speaking, these conditions identify the geodesics that pass through the locations of the monopoles.
As an example, the spectral curve of a 1-monopole is given by
\be
2\eta\zeta(X_1-iX_2)+\zeta(1+|{\bf X}|^2-2X_3)-\eta(1+|{\bf X}|^2+2X_3)-2(X_1+iX_2)=0,
\label{star}
\ee
where ${\bf X}=(X_1,X_2,X_3)$ is the point inside the unit ball at which the monopole is located (given by the vanishing of the Higgs field). This spectral curve gives all the geodesics that pass through the point ${\bf X}.$

A monopole is parity inversion symmetric if it invariant under changing the sign of all the Cartesian coordinates together with the sign of the Higgs field.
The spectral curve of a parity inversion symmetric $N$-monopole satisfies
\be
S(\eta,\zeta)=\eta^N\zeta^N\overline{S(-\bar\eta^{-1},-\bar \zeta^{-1})}.
\label{invsym}
\ee
For example, the spectral curve (\ref{star}) of the 1-monopole is parity inversion symmetric only if the monopole is located at the origin, ${\bf X}={\bf 0}$.

The reason for using the spectral curve description of monopoles in this paper is that there is an integral formula for the boundary metric in terms of the spectral curve \cite{MNS}. Let the most general $N$-monopole spectral curve $S(\eta,\zeta)$ be given in terms of real parameters $\alpha_1,\alpha_2,...,\alpha_{4N-1}$, that are coordinates on the $(4N-1)$-dimensional moduli space. A real function $h(z,\bar z)$ is constructed by evaluating the spectral curve on the anti-diagonal $\bar\Delta,$ via the definition
\be
h(z,\bar z)=\bar z^N S(-\bar z^{-1},z).
\ee
The formula for the boundary metric $ds^2=g_{ij}d\alpha_i d\alpha_j$ is \cite{MNS}
\be
g_{ij}=\frac{3}{4\pi}\int \Re\bigg\{
\bigg(\frac{\partial}{\partial \alpha_i}\frac{\partial }{\partial z}\log h\bigg)
\bigg(\frac{\partial}{\partial \alpha_j}\frac{\partial }{\partial \bar z}\log h\bigg)
\bigg\}
\, idzd\bar z,
\label{bmetric}
\ee
where the integration is over the whole plane and $\Re$ denotes the real part.
The normalization factor has been fixed by the requirement that the metric of a single monopole restricted to a radial half-line from the origin is $d\rho^2$, where $\rho$ is the geodesic distance from the origin \cite{BCS}. To date, the only calculation of a boundary metric for hyperbolic monopoles is the simplest case of the three-dimensional 1-monopole moduli space, where the result is 3-dimensional hyperbolic space \cite{BA}. Interpreting the geodesics of this manifold in terms of monopole dynamics provides a description of the free motion of a single monopole. In the next section the first example relevant for hyperbolic monopole scattering is constructed, by calculating the boundary metric on the hyperbolic analogue of the 4-dimensional Atiyah-Hitchin manifold for charge two monopoles with parity inversion symmetry.  

\section{The metric}\quad
The identification of a hyperbolic $N$-monopole with a circle-invariant Yang-Mills $N$-instanton allows the ADHM construction of instantons \cite{ADHM} to be adapted to a construction for hyperbolic monopoles \cite{MS}. Within this framework, the data for a hyperbolic $N$-monopole consists of an $N$-component row vector $L$ of quaternions, together with a triplet $M_1,M_2,M_3$ of real $N\times N$ symmetric matrices. These are assembled to form the $(N+1)\times N$ matrix of quaternions
\be
\widehat M=\begin{pmatrix} L \\ iM_1+jM_2+kM_3 \end{pmatrix},
\ee
that is required to satisfy the condition that
$\widehat M^\dagger \widehat M$ is the identity matrix, together with the constraint that $L(iM_1+jM_2+kM_3)=\mu L$, for some pure quaternion $\mu.$ Here $\dagger$ denotes the quaternionic conjugate transpose.

The 1-parameter family of 2-monopole data of interest for the current application is given by
\be
M_1=\begin{pmatrix} 0&0\\0&0 \end{pmatrix}, \
M_2=\frac{1}{2}(1-a)\begin{pmatrix} 0&1\\1&0 \end{pmatrix}, \
M_3=\frac{1}{2}(1+a)\begin{pmatrix} 1&0\\0&-1 \end{pmatrix}, \
L=\sqrt{\frac{1-a^2}{2}}(1,-i),
\label{data0}
\ee
with parameter $a\in[0,1)$. It is easy to verify that this data satisfies the two conditions stated above, with $\mu=ak.$
Applying the adapted ADHM construction to this data produces an explicit formula \cite{MS} for the Higgs field of a pair of monopoles, one on the positive $x_3$-axis and one on the negative $x_3$-axis, each with geodesic distance $\rho$ to the origin given by
\be
\rho=2\tanh^{-1}\bigg(\sqrt{\frac{3-a^2-\sqrt{(1-a^2)(9-a^2)}}{2a}}\bigg).
\label{distance}
\ee
The above is true if $a>0$, so that the two distinct monopole positions can be identified with the two distinct zeros of the Higgs field. If $a=0$ then $\rho=0$ and the Higgs field of the monopole vanishes only at the origin. In this case the 2-monopole is axially symmetric about the $x_1$-axis. Copying the notation from Euclidean 2-monopoles \cite{AH}, each parity inversion symmetric 2-monopole is characterized by three orthogonal, unoriented lines passing through the origin, that may be viewed as the body-fixed axes $e_1,e_2,e_3$. The $e_3$-axis is the line passing through both zeros of the Higgs field, so the $x_3$-axis for the above data. The $e_1$-axis is the symmetry axis when the two zeros of the Higgs field coalesce, being the $x_1$-axis for this data.

The formula for the spectral curve in terms of adapted ADHM data is \cite{Su2} 
\be
S(\eta,\zeta)=\mbox{det}\big(\eta\zeta(M_1-iM_2)+\zeta(1-M_3)-\eta(1+M_3)-(M_1+iM_2)\big).
\label{sc}
\ee
Inserting the data (\ref{data0}) gives
\be
\frac{1}{4} \left( 1-a \right) ^{2}\left( {\eta}^{2}{\zeta}^{2}+1 \right) +\frac{1}{4} (3+a)(1-a)
 \left( {\eta}^{2}+{\zeta}^{2} \right) 
-2  \left( 1+a \right) \eta \zeta=0.
\ee
It is easily checked that this spectral curve satisfies the condition (\ref{invsym}) to be parity inversion symmetric. 
If $a=0$ then the axial symmetry of the curve around the $x_1$-axis is apparent from the invariance under rotations 
\be
(\eta,\zeta)\mapsto \bigg(
\frac{\cos(\gamma/2)\eta+i\sin(\gamma/2)}{i\sin(\gamma/2)\eta+\cos(\gamma/2)},
\frac{\cos(\gamma/2)\zeta+i\sin(\gamma/2)}{i\sin(\gamma/2)\zeta+\cos(\gamma/2)},
\bigg),
\ee
for an arbitrary angle $\gamma.$
In the limit $a\to 1$ the curve degenerates to a product $\eta\zeta=0,$ corresponding to a pair of single monopoles on the positive and negative $x_3$-axis, at infinite geodesic distance from the origin, as obtained by inserting ${\bf X}=(0,0,\pm 1)$ into the formula (\ref{star}) for the 1-monopole spectral curve.

The full 4-dimensional family of parity inversion symmetric 2-monopoles is obtained by applying an arbitrary $SO(3)$ rotation to the data (\ref{data0}). In terms of Euler angles $\theta,\phi,\psi$ define the $SO(3)$ matrix
\be
Q=
\left[ \begin {array}{ccc} \cos \psi \cos \theta \cos \phi -\sin\psi \sin \phi  &\sin \psi \cos \theta \cos \phi +\cos \psi \sin \phi &\sin \theta \cos \phi \\ \noalign{\medskip}-\cos \psi \cos \theta \sin \phi -\sin \psi \cos \phi &-\sin
\psi \cos \theta \sin \phi +\cos \psi \cos \phi &-\sin \theta \sin \phi \\ \noalign{\medskip}-\cos \psi \sin \theta &-\sin \psi \sin \theta &\cos \theta  \end {array} \right] 
\ee
and obtain the required data by applying the rotation
\bea
\begin{pmatrix}M_1\\M_2\\M_3\end{pmatrix}
  \mapsto
  Q
  \begin{pmatrix}M_1\\M_2\\M_3\end{pmatrix}, \quad
L\mapsto     
\sqrt{\frac{1-a^2}{2}}\left(1,
k\cos\psi\sin\theta+(j\sin\psi-i\cos\psi\cos\theta)e^{k\phi})\right)
.
  \eea
  Using the formula (\ref{sc}) gives the spectral curve of the 4-dimensional moduli space of parity inversion symmetric 2-monopoles, with coefficients
  \bea
  &c_{00}=\overline{c_{22}}=
-\frac{1}{4} {{\rm e}^{-2 i\phi}} \left(  \left( 1-a \right) ^{2} \left( 
i\cos  \theta \, \sin  2 \psi  -  \cos^2\psi \cos^2 \theta 
 + \sin^2 \psi  \right) +
4 a \sin^2\theta  \right) \nonumber
\\
&c_{01}=c_{10}=-\overline{c_{12}}=-\overline{c_{21}}=
\frac{1}{4}{{\rm e}^{-i\phi}}\sin\theta  \left(  \left( 1-a
 \right) ^{2} \left( i\sin 2\psi  -2\cos\theta\cos^2 \psi \right) 
 -8a\cos\theta \right)\nonumber\\
 &c_{02}=c_{20}=
 \frac{1}{4} \left( 1-a \right) ^{2} \sin^2 \theta \cos^2 \psi
 +\frac{1}{4}\left( 3-2a\cos 2\theta -{a}^{2}\right) \nonumber \\
 &c_{11}=
 \left( 1-a \right) ^{2} \sin^2 \theta \cos^2 \psi -2a\cos 2\theta -2.
 \label{coeffs}
 \eea
 Note that for $N=2$ the symmetry of the coefficients, $c_{ij}=c_{ji}$, follows directly from the combined application of parity inversion symmetry (\ref{invsym}) and the reality condition (\ref{reality}).

As with the Atiyah-Hitchin manifold, the $SO(3)$ and reflection symmetries imply that the metric has the form
\be
ds^2 = \Lambda_0^2  \, da^2 + \Lambda_1^2 \, \sigma_1^2 + \Lambda_2^2 \, \sigma^2_2 + \Lambda_3^2\, \sigma^2_3,
\label{so3sym}
\ee
where $\Lambda_i^2(a)$ are functions of $a$ only, and
$\sigma_i$ are the standard 1-forms on $SO(3)$,
\begin{eqnarray}
\sigma_1 &=& -\sin \psi \, d \theta + \cos \psi \sin \theta \, d\phi\\
\sigma_2 &=& \cos \psi \, d \theta + \sin\psi \sin \theta \, d\phi\\
\sigma_3 &=& d \psi  + \cos  \theta \, d\phi\,.
\end{eqnarray}
For $i=1,2,3$ the function $\Lambda_i^2(a)$ is the moment of inertia about the body-fixed axis $e_i$.
The spectral curve coefficients (\ref{coeffs}), together with the integral formula (\ref{bmetric}), provide all the ingredients required for a computation of the metric on the 4-dimensional moduli space of parity inversion symmetric  charge two hyperbolic monopoles.

Any convenient orientation may be chosen for the calculation of the metric functions. In particular, once any differentiation with respect to Euler angles required in (\ref{bmetric}) has been performed, then the Euler angles may be set to values that will facilitate the calculation of the integral. A particularly convenient choice is $\theta=\pi/2,\, \phi=\psi=0$, so given any function of the Euler angles, say $f(\theta,\phi,\psi)$, it is helpful to introduce the notation $f_\bullet=f(\pi/2,0,0).$

\begin{figure}[ht]\begin{center}
    \hbox{
      \includegraphics[width=0.51\columnwidth]{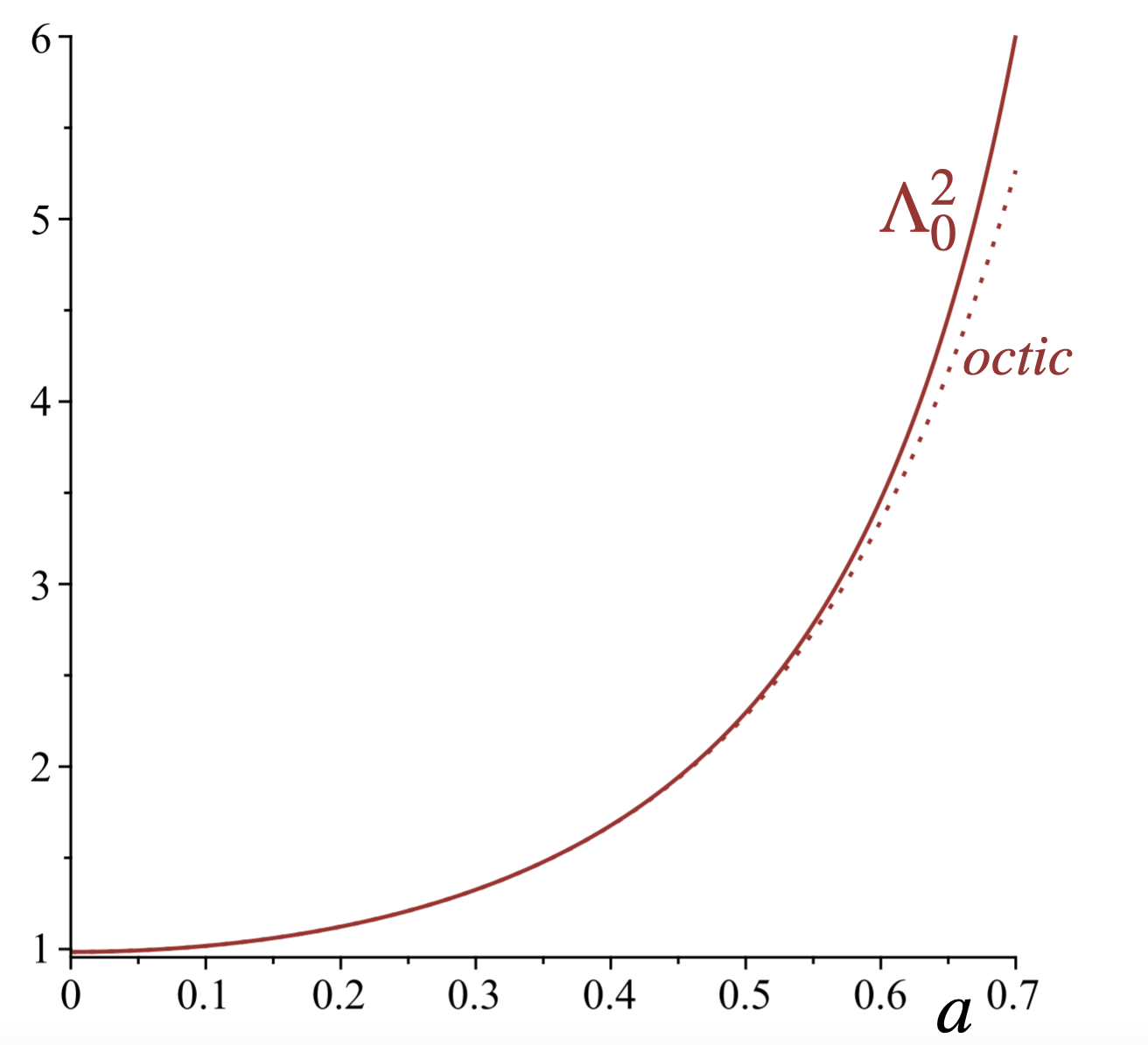}
      \includegraphics[width=0.49\columnwidth]{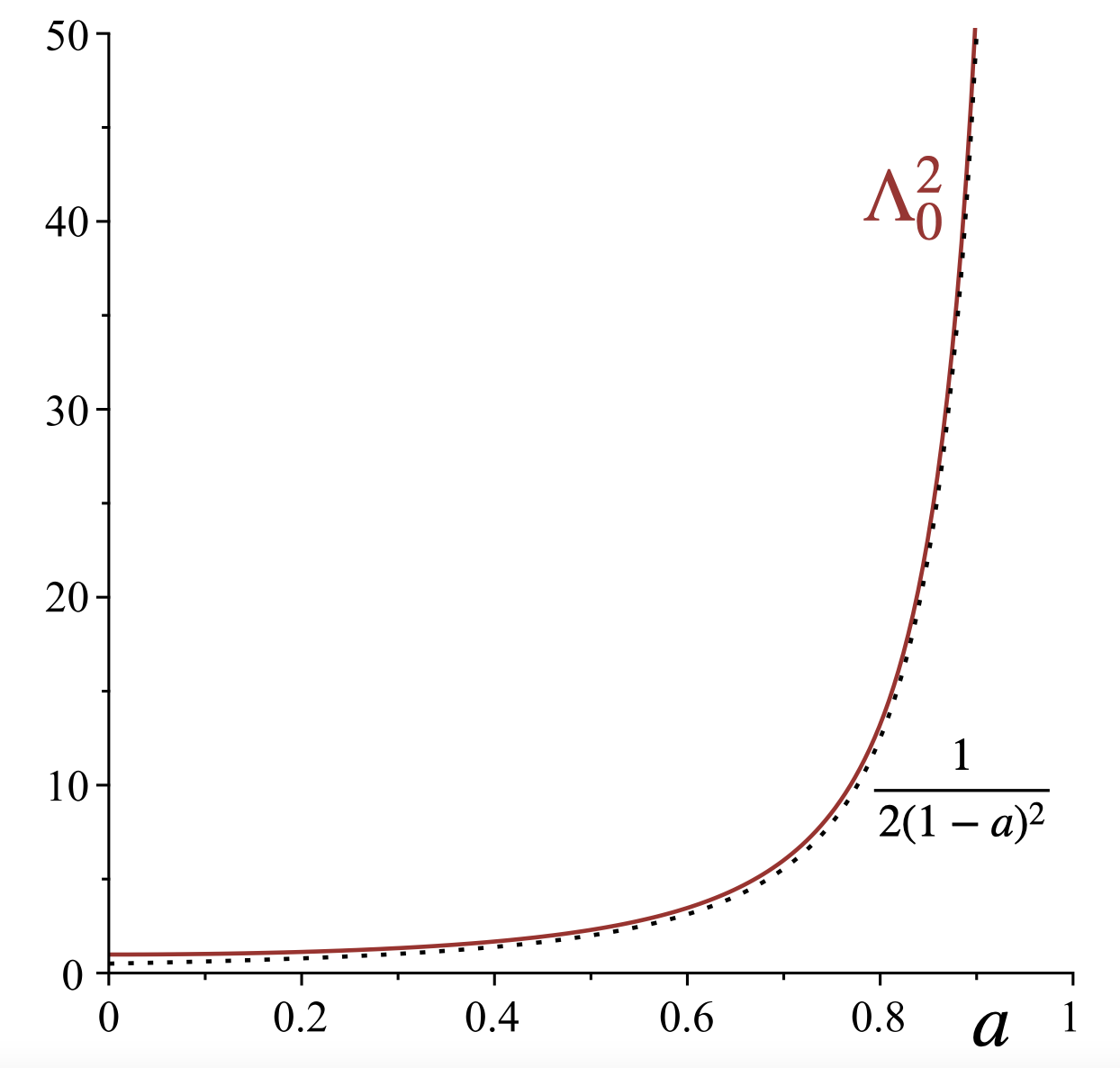}
      }
    \caption{The metric function $\Lambda_0^2(a)$ with octic approximation (left), and asymptotics (right).}
      \label{fig:Lam0}\end{center}\end{figure}

The calculation of the metric function $\Lambda_0^2(a)$ requires no differentiation with respect to Euler angles, so they may be fixed immediately to simplify the spectral curve to
\be
S(\eta,\zeta)_\bullet=-a(\eta^2\zeta^2+1)+\eta^2+\zeta^2-(1-a^2)\eta\zeta.
\ee
This gives the real function
\be
h(z,\bar z)_\bullet=1+|z|^4+(1-a^2)|z|^2-a(z^2+\bar z^2),
\ee
and hence the integral formula
\bea
\Lambda_0^2(a)&=&\frac{3}{4\pi}\int \bigg|
\frac{\partial }{\partial z}\frac{\partial}{\partial a}\log h\bigg|^2_\bullet
\,
idzd\bar z. 
\\
&=&
\frac{3}{4\pi}\int 
\frac{\big|a^2\bar z(\bar z^2-z^2)+2a\bar z(|z|^4-1)+\bar z^3(2|z|^2+1)-z(|z|^2+2)\big|^2}
{\big(1+|z|^4+(1-a^2)|z|^2-a(z^2+\bar z^2)\big)^4}
idzd\bar z. \qquad
\label{lam0z}
\eea
By making the substitution $z=\sqrt{p}e^{i\chi},$ the integration over $\chi$ may be performed by using the results that the integrals
\be
I_n(b)=\int_0^{2\pi} \frac{\cos^n\chi}{(1-b\cos\chi)^4} \, d\chi,
\ee
with parameter $b\in[0,1),$ are given by
\be
I_0(b)=\frac{\pi(2+3b^2)}{(1-b^2)^{7/2}},\quad
I_1(b)=\frac{\pi b(4+b^2)}{(1-b^2)^{7/2}},\quad
I_2(b)=\frac{\pi (1+4b^2)}{(1-b^2)^{7/2}}.\quad
I_3(b)=\frac{\pi b(3+2b^2)}{(1-b^2)^{7/2}}.\quad
\ee
Applying these identities to (\ref{lam0z}) yields
\begin{dmath}
  \Lambda_0^2(a)=\int_0^\infty \bigg(
\left( 2{a}^{2}+2 \right)  \left( {p}^{10}+1 \right) + \left( -6{
a}^{4}+18{a}^{2}+8 \right)  \left( {p}^{9}+p \right) + \left( 6{a}
^{6}-31{a}^{4}+42{a}^{2}+19 \right)  \left( {p}^{8}+{p}^{2}
 \right) + \left( -2{a}^{8}+11{a}^{6}-33{a}^{4}+9{a}^{2}+31
 \right)  \left( {p}^{7}+{p}^{3} \right) + \left( {a}^{8}+2{a}^{6}+9
{a}^{4}-92{a}^{2}+40 \right)  \left( {p}^{6}+{p}^{4} \right) +
 \left( -{a}^{10}+9{a}^{8}-8{a}^{6}+32{a}^{4}-147{a}^{2}+43
 \right) {p}^{5}
 \bigg)\\
 \bigg(  \left( 1+{p}^{2}+ \left( -{a}^{2}+2a+1 \right) p \right) 
 \left( 1+{p}^{2}+ \left( -{a}^{2}-2a+1 \right) p \right)  \bigg) ^
      {-7/2} 3p\,dp,
\label{Lam0}\end{dmath}
which is an elliptic integral, as the integrand is a product of a rational function of $p$ and the square root of a quartic in $p$. This elliptic integral representation for  $\Lambda_0^2(a)$ is perfectly acceptable for both numerical evaluation and the calculation of a series expansion about $a=0$, to any desired order. For example, the series to octic order is
\begin{dmath} 
 \Lambda_0^2(a)=
3-{\frac {10\sqrt {3}\pi}{27}}+ \left( 7-{\frac {56\sqrt {3}\pi}{
81}} \right) {a}^{2}+ \left( 11-{\frac {758\sqrt {3}\pi}{729}}
 \right) {a}^{4}+ \left( 15-{\frac {9176\sqrt {3}\pi}{6561}}
 \right) {a}^{6}+ \left( 19-{\frac {34670\sqrt {3}\pi}{19683}}
 \right) {a}^{8}
 +...
 \label{octicLam0}
\end{dmath}
Fig.\ref{fig:Lam0} displays a plot of $\Lambda_0^2(a).$ In the figure on the left the octic approximation (\ref{octicLam0}) is shown as the dashed line, and is found to be an excellent approximation, even up to reasonably large monopole separations. The asymptotic behaviour for large monopole separation ($a\sim 1$) is shown in the figure on the right. The limit is
\be
\lim_{a\to 1}(1-a)^2\Lambda_0^2(a)=\frac{1}{2},
\ee
so that for $a\sim 1$ the limiting behaviour, together with (\ref{distance}), gives $\Lambda_0^2 da^2 \sim 2d\rho^2,$ associated with a pair of monopoles, each with geodesic distance $\rho$ to the origin.

With some effort, the elliptic integral (\ref{Lam0}) can be written in terms of standard elliptic integrals. The required change of variable is
\be
p=\frac{1+a+u\sqrt{(3+a)(1-a)}}{1+a-u\sqrt{(3+a)(1-a)}},
\label{p2u}
\ee
and leads to the formula
  \bea
   \Lambda_0^2(a)=
        2{a}^{-2} \left( a+3 \right) ^{-2} \left( 1-a \right) ^{-2}\left( 3-a \right) ^{-5/2} \left( 1+a \right) ^{-7/2} \nonumber \\ 
  \bigg( -24 { F(\varphi,k)} a  \left( a+3 \right)  \left( 1-a
  \right) ^{2} \left( {a}^{4}-2 {a}^{3}-4 {a}^{2}+18 a+3 \right) \nonumber \\
  +3 \Pi(1,\varphi,k)   \left( 1-a \right) ^{3} \left( a+3 \right)  \left( {a}^{6
  }-27 {a}^{4}+99 {a}^{2}-9 \right) \nonumber \\
  +\sqrt {3-a} \left( 1+a \right) ^{
3/2} \left( {a}^{2}+3 \right)  \left( {a}^{6}-9 {a}^{4}+63 {a}^{2}+9
  \right) \bigg).
  \eea
Here $F(\varphi,k)$ and $\Pi(1,\varphi,k)$ are the standard elliptic integrals of the first and third kind, defined by
    \be
  F(\varphi,k)=\int_0^{\sin\varphi} \frac{du}{\sqrt{(1-u^2)(1-k^2u^2)}},
  \quad
  \Pi(1,\varphi,k)=
  \int_0^{\sin\varphi} \frac{du}{\sqrt{(1-u^2)^3(1-k^2u^2)}},
    \ee
 with amplitude $\varphi$ and elliptic modulus $k$ given by   
 \be
 \varphi=\sin^{-1}((1+a)/2), \qquad\quad  k=\sqrt{\frac{16a}{(1+a)^3(3-a)}}.
 \label{ampmod}
 \ee
A similar calculation for $\Lambda_3^2(a)$, using the integral formula
\be
\Lambda_3^2(a)=\frac{3}{4\pi}\int \bigg|
\frac{\partial }{\partial z}\frac{\partial}{\partial \psi}\log h\bigg|^2_\bullet
\,
idzd\bar z,
\ee
produces 
\begin{dmath}
  \Lambda_3^2(a)=\frac{3}{8}\int_0^\infty 
\bigg({p}^{10}+1-2    \left( {a}^{2}+a+1 \right)\left( {p}^{9}+p \right) +
   \left( 2 {a}^{4}+12 {a}^{2}+8 a-1
 \right)\left( {p}^{8}+{p}^{2} \right) +2  \left( -{a}^{6}+{a}^{5}+{a}^{4}-10 {a}^{3}-15 {a}^{2}+
5 a+7 \right)  \left( {p}^{7}+{p}^{3} \right) + \left( {a}^{8}-8 {a}
^{6}+52 {a}^{4}-100 {a}^{2}-8 a+41 \right)  \left( {p}^{6}+{p}^{4}
 \right) +4  \left( -5 {a}^{6}-{a}^{5}+31 {a}^{4}+10 {a}^{3}-30 {
   a}^{2}-4 a+14 \right) {p}^{5} \bigg) \\
 \bigg( 1+{p}^{2}+ \left( -{a}^{2}+2 a+1 \right) p \bigg) ^{-7/2}
 \bigg( 1+{p}^{2}+ \left( -{a}^{2}-2 a+1 \right) p \bigg) ^{-5/2}
 (1-a)^4\,dp,
 \end{dmath}
with an octic approximation
\begin{dmath} 
  \Lambda_3^2(a)=\frac{2}{9}\sqrt {3}\pi-1+ \left( \frac{3}{2}-\frac{4}{9}\sqrt {3}\pi \right) a+ \left( -\frac{7}{2}+{\frac {26\sqrt {3}\pi}{27}} \right) {a}^{2}+ \left( 6-{\frac {40
\sqrt {3}\pi}{27}} \right) {a}^{3}+ \left( -{\frac{26}{3}}+{\frac {
478\sqrt {3}\pi}{243}} \right) {a}^{4}+ \left( {\frac{34}{3}}-{
\frac {596\sqrt {3}\pi}{243}} \right) {a}^{5}+ \left( -14+{\frac {
6434\sqrt {3}\pi}{2187}} \right) {a}^{6}+ \left( {\frac{50}{3}}-{
\frac {7504\sqrt {3}\pi}{2187}} \right) {a}^{7}+ \left( {\frac {
25726\sqrt {3}\pi}{6561}}-{\frac{58}{3}} \right) {a}^{8}
+...
\end{dmath}
Again the transformation (\ref{p2u}) allows the result to be written in terms of standard elliptic integrals, yielding
\be
\Lambda_3^2(a)=\frac{(1-a)^2\left(8(1-a)F(\varphi,k)-(1-a)(3+a)(3+a^2)\Pi(1,\varphi,k)+(1+a)^{5/2}\sqrt{3-a}\,\right)}
       {2(1+a)^{7/2}\sqrt{3-a}},
       \label{Lam3}
\ee
with the same amplitude and elliptic modulus as earlier, given by (\ref{ampmod}).

In the limit of infinite monopole separation there is axial symmetry around the $e_3$-axis and 
\be
\lim_{a\to 1}\Lambda_3^2(a)=0.
\ee
In the Euclidean case the analogous moment of inertia tends to a non-zero constant in the limit of infinite monopole separation, as motion in the relative phase turns the monopoles into dyons with opposite electric charge, which contributes to the kinetic energy. In the hyperbolic case the boundary metric detects this contribution for any finite monopole separation, but not in the limit of infinite separation. This is because the boundary metric is obtained by a renormalization of the divergent kinetic energy induced from the field theory \cite{Su1}, and in the limit $a\to 1$ the associated tangent vector has finite length and therefore does not contribute to the divergence of the kinetic energy. This corresponds to the fact that a single hyperbolic monopole can be promoted to a finite energy hyperbolic dyon.

Performing the calculation
\be
\Lambda_2^2(a)=\frac{3}{4\pi}\int \bigg|
\frac{\partial }{\partial z}\frac{\partial}{\partial \theta}\log h\bigg|^2_\bullet
\,
idzd\bar z,
\ee
reveals that $\Lambda_2^2(a)=\Lambda_3^2(-a)$. This is a reflection of the fact that making the replacement $a\mapsto -a$ in the spectral curve swaps the $e_2$ and $e_3$ axes. In particular, the axial symmetry at $a=0$ requires that $\Lambda_2^2(0)=\Lambda_3^2(0)$. 

Using the relation $\Lambda_2^2(a)=\Lambda_3^2(-a)$ together with the formula (\ref{Lam3}) gives that
\be
\lim_{a\to 1}(1-a)\Lambda_2^2(a)=2.
\ee
Therefore in the limit of large monopole separations the asymptotic behaviour is
\be
\Lambda_2^2(a)\sim 2/(1-a), \qquad \mbox{for \ } \  a\sim 1.
\ee
The remaining component of the metric can be calculated using the formula
\be
\Lambda_1^2(a)=\frac{3}{4\pi}\int \bigg|
\frac{\partial }{\partial z}\frac{\partial}{\partial \phi}\log h\bigg|^2_\bullet
\,
idzd\bar z,
\ee
to give the elliptic integral representation
\begin{dmath} 
  \Lambda_1^2(a)=\int_0^\infty \bigg(
2 ({p}^{7}+ p)+ \left( -4 {a}^{2}+4 \right)  \left( {p}^{6}+{p}^{2}
 \right) + \left( 3 {a}^{4}-10 {a}^{2}+5 \right)  \left( {p}^{5}+{p}
^{3} \right) + \left( -{a}^{6}+7 {a}^{4}-11 {a}^{2}+5 \right) {p}^{4
}
 \bigg)
 \\
 \bigg( 1+{p}^{2}+ \left( -{a}^{2}+2 a+1 \right) p \bigg) ^{-5/2}
 \bigg( 1+{p}^{2}+ \left( -{a}^{2}-2 a+1 \right) p \bigg) ^{-5/2}
 12a^2\,dp
\end{dmath}  
with octic expansion
\begin{dmath} 
  \Lambda_1^2(a)=
  \left(12 -{\frac {40\sqrt {3}\pi}{27}} \right) {a}^{2}+ \left({\frac{68}{3}} -{
\frac {832\sqrt {3}\pi}{243}} \right) {a}^{4}+
 \left({\frac{100}{3}} -{\frac {3928\sqrt {3}\pi}{729}} \right) {a}
^{6}+ \left(44 -{\frac {48224\sqrt {3}\pi}{6561}} \right) {a}^{8}
 +...
 \label{octicLam1}
\end{dmath}
Note that $\Lambda_1^2(0)=0,$ corresponding to the axial symmetry about the $e_1$ axis when $a=0.$
In terms of standard elliptic integrals the formula for $\Lambda_1^2(a)$ is
\begin{dmath}
  \Lambda_1^2(a)=2\bigg(
  \left( 3+a \right)  \left( 1-a \right) ^{2} \left( 8 a 
            \left( 1-3 a \right) F(\varphi,k)+ \left(3 +30 {a}^{2}-a^4 \right) \Pi(1,\varphi,k)\right)\\
                        -  \left( 1+a \right) ^{5/2}\sqrt {3-a} \left( 3-20 {a}^{2}+a^4 \right)\bigg)  \left( 1+a \right) ^{-7/2} \left( 3-a \right) ^{-3/
2} \left( 1-a \right)^{-1}  \left( 3+a \right)^{-1}.
\end{dmath}
Using the result
\be
\lim_{a\to 1}(1-a)\Lambda_1^2(a)=2,
\ee
confirms that in the limit of large monopole separations ($a\sim 1$) the asymptotic behaviour is
$\Lambda_1^2(a) \sim 2/(1-a) \sim \Lambda_2^2(a)$, reflecting the properties of point-like monopoles. 

\begin{figure}[ht]\begin{center}
    \hbox{
      \includegraphics[width=0.52\columnwidth]{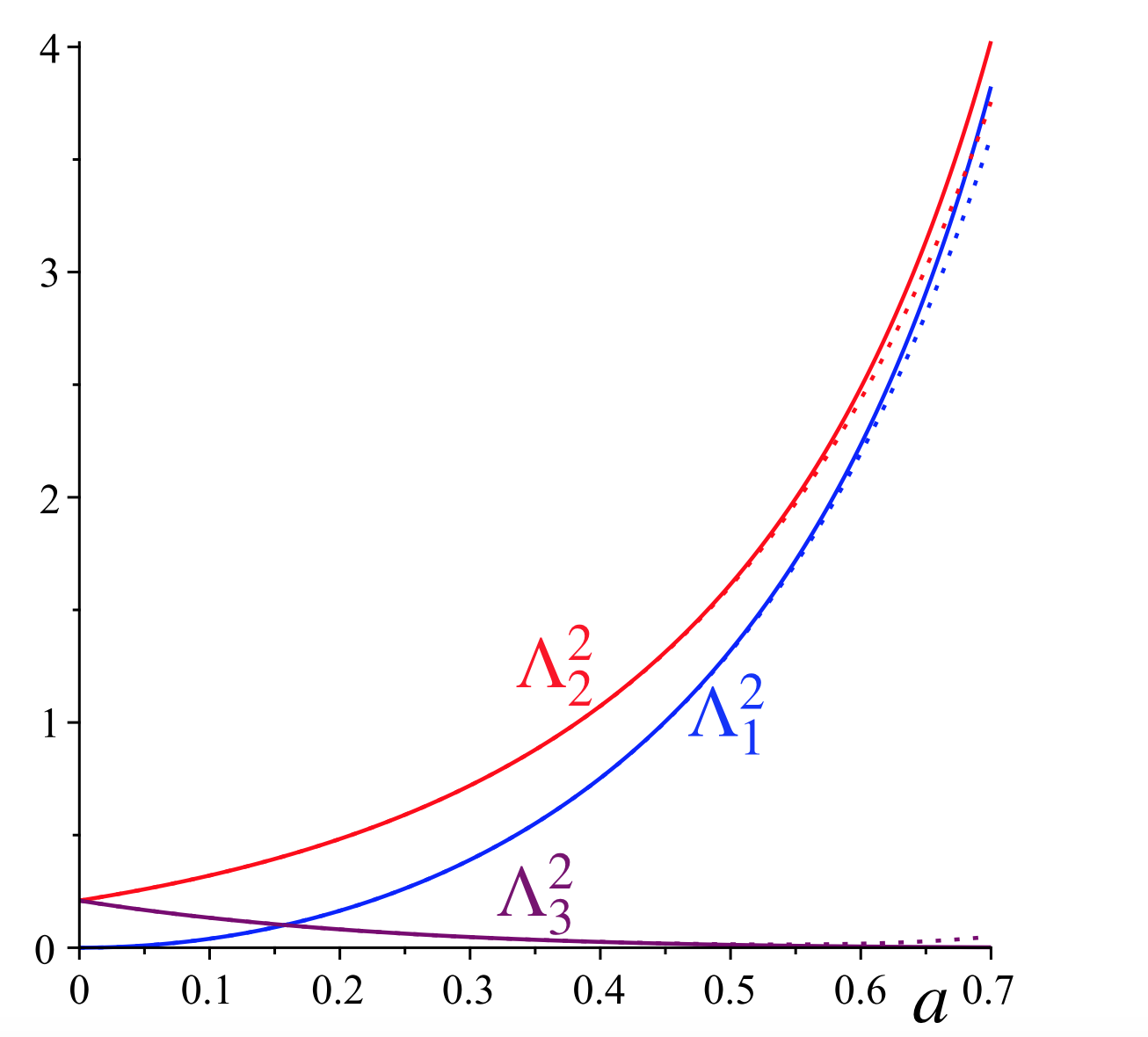}
      \includegraphics[width=0.47\columnwidth]{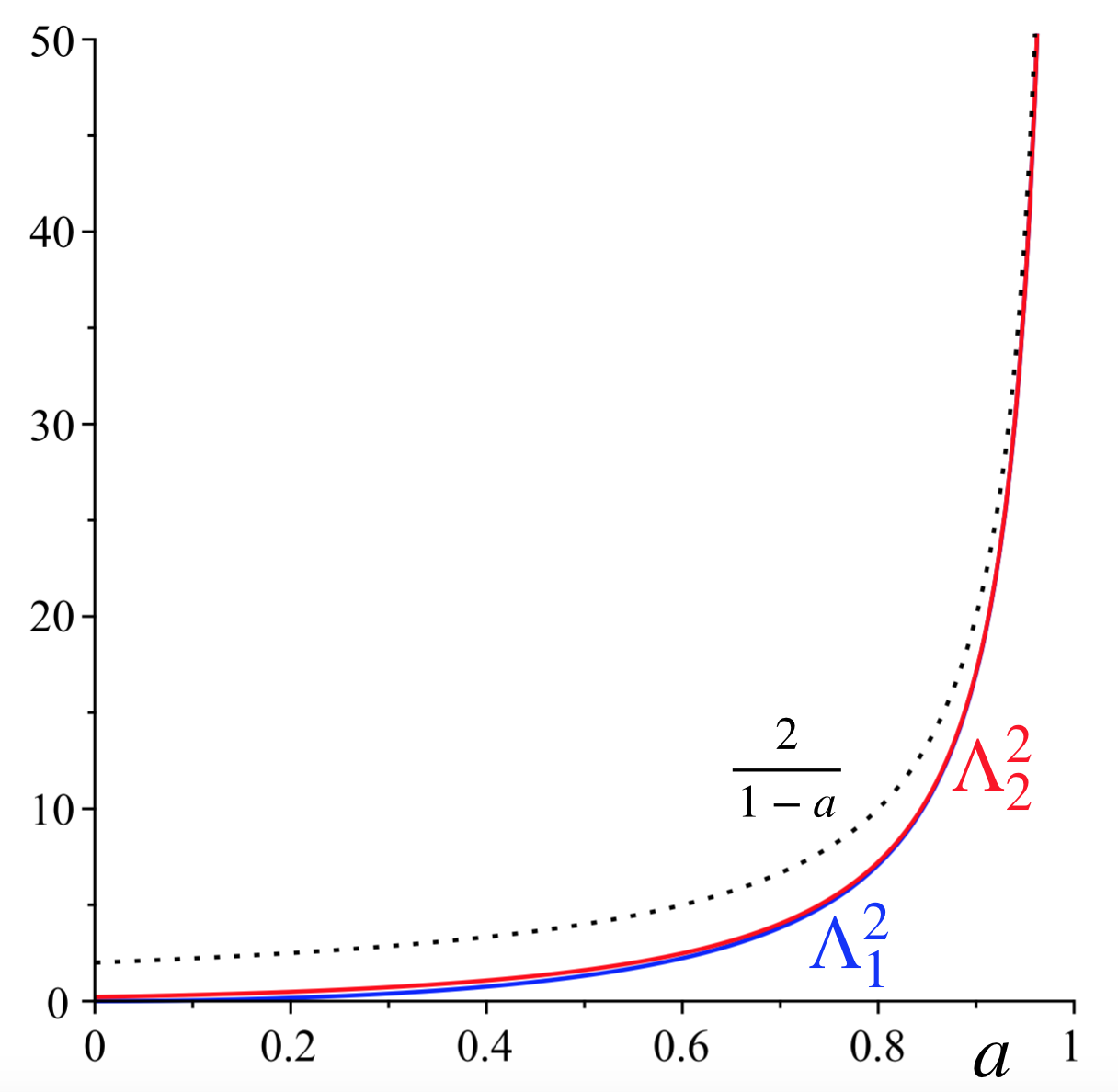}
      }
\caption{The metric functions $\Lambda_1^2(a),\Lambda_2^2(a),\Lambda_3^2(a)$ with octic approximations (left), and asymptotics (right).}
\label{fig:Lam123}\end{center}\end{figure}
All three metric functions $\Lambda_1^2(a),\Lambda_2^2(a),\Lambda_3^2(a)$ are displayed in Fig.\ref{fig:Lam123}. In the figure on the left the octic expansions are shown as the dashed lines, and again provide an excellent approximation, with barely discernible errors at this scale. In the figure on the right the dashed line shows the common asymptotic behaviour of $\Lambda_1^2(a)$ and $\Lambda_2^2(a).$ These plots demonstrate a qualitative agreement with the corresponding metric functions for the Atiyah-Hitchin manifold \cite{AH}, modulo the asymptotic value of $\Lambda_3^2$, as discussed earlier. This justifies the interpretation of this 4-manifold as a hyperbolic analogue of the Atiyah-Hitchin manifold and confirms the similarity of Euclidean and hyperbolic monopole scattering within the geodesic approximation.

\section{A rounded cone and a tapered trumpet}\quad
The Atiyah-Hitchin manifold contains totally geodesic 2-dimensional submanifolds, known as the rounded cone and the trumpet, that are surfaces of revolution. In this section their hyperbolic analogues are presented, together with a description in terms of rational maps.

Fixing the principal axis $e_1$ to be the $x_3$-axis corresponds to setting $\theta=\pi/2$ and $\psi=0$ and is a totally geodesic 2-dimensional submanifold $\Sigma_{\rm cone}$. The associated spectral curve is
\be
a(e^{2i\phi}\eta^2\zeta^2+e^{-2i\phi})+\eta^2+\zeta^2-(1-a^2)\eta\zeta=0,
\ee
which shows that $\phi$ has period $\pi$ for this submanifold, rather than period $2\pi$ in the full manifold. It is therefore convenient to introduce the coordinate $\widetilde\phi=2\phi\,$ on $\Sigma_{\rm cone}$, with period $2\pi.$ The metric induced from (\ref{so3sym}) is then
\be
ds^2=\Lambda_0^2\, da^2+\frac{1}{4}\Lambda_1^2\, d\widetilde\phi^2,
\ee
which is a rounded cone. Using the series expansions (\ref{octicLam0}) and (\ref{octicLam1}) gives the leading order behaviour at the vertex $a=0$,
\be
ds^2=\frac{1}{27}(81 -{10\sqrt {3}\pi})
( da^2+a^2\, d\widetilde\phi^2)+...
\ee
confirming that the cone is smooth at the vertex. This vertex corresponds to the axially symmetric 2-monopole, with the $x_3$-axis being the axis of symmetry.
The generating geodesics of the rounded cone pass through the vertex and describe the right angle scattering of monopoles in a head-on collision, with $\widetilde\phi$ constant during the evolution, except for a jump by $\pi$ as the geodesic passes over the cone. This gives a jump of $\pi/2$ in $\phi$ and hence the monopoles separate at right angles to the direction of their approach.

The rational map of a hyperbolic $N$-monopole is a degree $N$ based rational map between Riemann spheres, ${\cal R}(z)$, with the base point condition ${\cal R}(\infty)=0.$ The rational map may be viewed as scattering data in the background of the hyperbolic $N$-monopole, along the geodesic associated with the point $(\eta,\zeta)=(0,z)$ in mini-twistor space. The scattering data, and hence the rational map, is defined only up to the equivalence relation of multiplication by a phase. There is a bijective correspondence between these equivalence classes of rational maps and the $(4N-1)$-dimensional moduli space of hyperbolic $N$-monopoles \cite{At,At2}.

To obtain the rational map from the ADHM data of the hyperbolic monopole, introduce the rank one Hermitian matrix
\be
H=
(1-M_3)^{-1/2}\bigg(1+M_3-(M_1-iM_2)(1-M_3)^{-1}(M_1+iM_2)\bigg)(1-M_3)^{-1/2},
\label{H}
\ee
and let $v$ be its unit-length eigenvector with non-zero eigenvalue $\lambda.$ The formula for the rational map is
\cite{Su2}
\be
   {\cal R}(z)=\lambda v^\dagger\bigg(z-(1-M_3)^{-1/2}(M_1+iM_2)(1-M_3)^{-1/2}\bigg)^{-1}\bar v.
   \label{rat}
   \ee
   Applying this to $\Sigma_{\rm cone}$ provides its rational map description
   \be
      {\cal R}(z)=\frac{1-a^2}{z^2-ae^{-i\widetilde\phi}}
      =\frac{1-|c|^2}{z^2-c},
            \ee
            where the complex coordinate $c=ae^{-i\widetilde\phi}$, that lies inside the unit disc, has been introduced. Restricting $c$ to the real interval $(-1,1)$ is a geodesic that describes right angle scattering, with an exchange of the $e_2$ and $e_3$ axes as $c$ passes through zero.

To construct a second 2-dimensional geodesic submanifold, begin by setting $\theta=\phi=0$ to fix the principal axis $e_3$ to be the $x_3$-axis. This defines a 2-dimensional submanifold $\Sigma_{\rm trumpet}^+$ with spectral curve 
 \be
\frac{1}{4}\left( 1-a \right) ^{2}\left( e^{i\widetilde\psi}{\eta}^{2}{\zeta}^{2}+e^{-i\widetilde\psi} \right)+\frac{1}{4}(3+a)(1-a)
 \left( {\eta}^{2}+{\zeta}^{2} \right)  
 -2  \left( 1+a \right) \eta \zeta=0,
 \label{sctrumpet1}
\ee  
where $\widetilde\psi=2\psi$ has period $2\pi.$ The metric on $\Sigma_{\rm trumpet}^+$ is 
 \be
 ds^2=\Lambda_0^2\, da^2+\frac{1}{4}\Lambda_3^2\, d\widetilde\psi^2.
 \label{metrictrumpet1}
\ee
This surface of revolution is a tapered cylinder with the boundary at $a=0$ given by a circle of radius $\frac{1}{6}\sqrt{2\sqrt{3}\pi-9}$. The radius of the circular cross-section decreases as $a$ increases and tends to zero as $a\to 1.$
The rational map associated with  $\Sigma_{\rm trumpet}^+$ is
\be
   {\cal R}(z)=\frac{8(1+a)z}{(1-a)(3+a)z^2+(1-a)^2e^{-i\widetilde\psi}}.
   \label{rattrumpet1}
   \ee
 Setting $\theta=\psi=\pi/2\,$ fixes the principal axis $e_2$ to be the $x_3$-axis and defines another 2-dimensional submanifold $\Sigma_{\rm trumpet}^-$ with spectral curve 
 \be
\frac{1}{4}\left( 1+a \right) ^{2}\left( e^{i\widetilde\psi}{\eta}^{2}{\zeta}^{2}+e^{-i\widetilde\psi} \right)+\frac{1}{4}(3-a)(1+a)
 \left( {\eta}^{2}+{\zeta}^{2} \right)  
 -2  \left( 1-a \right) \eta \zeta=0,
 \label{sctrumpet2}
\ee  
where now $\widetilde\psi=2\phi+\pi$, which again has period $2\pi.$ The metric on $\Sigma_{\rm trumpet}^-$ is 
 \be
 ds^2=\Lambda_0^2\, da^2+\frac{1}{4}\Lambda_2^2\, d\widetilde\psi^2.
 \label{metrictrumpet2}
\ee
This surface of revolution also has a boundary at $a=0$ given by a circle of radius $\frac{1}{6}\sqrt{2\sqrt{3}\pi-9}$. This time the radius of the circular cross-section increases as $a$ increases and tends to infinity as $a\to 1.$
The rational map description of $\Sigma_{\rm trumpet}^-$ is
\be
   {\cal R}(z)=\frac{8(1-a)z}{(1+a)(3-a)z^2+(1+a)^2e^{-i\widetilde\psi}}.
   \label{rattrumpet2}
   \ee
   Joining the two surfaces $\Sigma_{\rm trumpet}^+$ and $\Sigma_{\rm trumpet}^-$ at their common boundary produces a totally geodesic 2-dimensional submanifold $\Sigma_{\rm trumpet}$, that is a tapered trumpet. The spectral curves (\ref{sctrumpet1}) and (\ref{sctrumpet2}) are related by the transformation $a\to-a$, as are the rational maps (\ref{rattrumpet1}) and  (\ref{rattrumpet2}). The spectral curve and rational map for $\Sigma_{\rm trumpet}$ are therefore given by (\ref{sctrumpet1}) and  (\ref{rattrumpet1}) with the range of $a$ extended to $(-1,1)$. The same is true for the metric (\ref{metrictrumpet1}) because of the relation $\Lambda_2^2(a)=\Lambda_3^2(-a).$ Defining the complex coordinate $C=e^{-i\widetilde\psi}(1-a)/(3+a)$, that lies inside the unit disc with the origin removed, allows the rational map for $\Sigma_{\rm trumpet}$ to be written as
   \be
      {\cal R}(z)=\frac{z(1-|C|^2)/|C|}{z^2+C}.
      \ee
      The narrow end of the trumpet corresponds to $C$ close to the origin, with the monopoles on the $x_3$-axis, and the broad end of the trumpet is $C$ close to the unit circle, with the monopoles in the plane $x_3=0$. On the circle $|C|=\frac{1}{3}$ the 2-monopole is axially symmetric, with the axis of symmetry lying in the plane $x_3=0$.

      The generating curves of the tapered trumpet are all the geodesics where the monopoles approach in the plane $x_3=0$, form an axially symmetric 2-monopole, and scatter at right angles to emerge along the $x_3$-axis.
Generic geodesics start and end at the broad end of the trumpet and describe monopoles that scatter in the plane $x_3=0$, but this can be via intermediate dyonic motion along the $x_3$-axis, with motion in the relative phase, for an arbitrarily long period of time.
      Note that dyonic motion in the narrow end of the trumpet must eventually return to the broad end because of the conserved quantity $\Lambda_3^2\frac{d \widetilde\psi}{dt}$ and the fact that $\Lambda_3^2\to 0$ as $a\to 1.$

\section{Conclusion}\quad
The qualitative similarity between the Atiyah-Hitchin manifold and its hyperbolic analogue indicates that hyperbolic monopole dynamics mirrors the Euclidean picture, within a geodesic description. However, despite the resemblance between these two manifolds, there is an important geometric distinction. The Atiyah-Hitchin manifold is hyperk\"ahler, unlike its hyperbolic relative, and this difference is most clearly reflected in the methods used to calculate the two metrics. The condition that a metric of the form (\ref{so3sym}) is hyperk\"ahler implies a set of first-order differential equations for the metric functions \cite{GP}. This system of differential equations can be solved in terms of complete elliptic integrals to determine the metric \cite{AH}. These differential equations do not hold for the metric functions of the hyperbolic analogue, removing this option for their construction and leaving brute-force integration as the only alternative. Fortunately, by tuning the curvature of hyperbolic space to a specific value (minus one in the units used here), the twistor data is sufficiently amenable that the integration required to calculate the boundary metric is tractable. The result is more complicated than the elegant Atiyah-Hitchin metric, being written in terms of incomplete rather than complete elliptic integrals, this being another reflection of its geometric inferiority. An attempt could be made to try and calculate the boundary metric for monopoles in hyperbolic space with an arbitrary constant negative curvature, but even if tractable, the result will be an even more unpleasant generalisation of the formulae presented here.

\end{document}